\documentclass[12pt,a4paper]{article}
\usepackage[utf8]{inputenc}
\usepackage[T1]{fontenc}
\usepackage{amsmath}
\usepackage{amsfonts}
\usepackage{amssymb}
\usepackage{graphicx}
\usepackage[caption=false]{subfig}
\usepackage[left=2.00cm, right=2.00cm, top=2.00cm]{geometry}

\usepackage{cite} 

\usepackage[
colorlinks=true,urlcolor=blue,anchorcolor=blue,citecolor=blue,filecolor=blue,linkcolor=blue,menucolor=blue,pagecolor=blue,linktocpage=true,pdfproducer=medialab,pdfa=true]{hyperref}

\usepackage{authblk}
\newcommand{\mailto}[1]{\href{mailto:#1}{#1}}

\title{\textbf{W-boson mass and electric dipole moments \newline from colour-octet scalars}\footnote{{(Partially) presented at the 30th International Symposium on Lepton Photon Interactions at High Energies, hosted by the University of Manchester, 10-14 January 2022.}}}

\author[a]{Hector Gisbert}
\author[b]{Víctor Miralles}
\author[c]{Joan Ruiz-Vidal\thanks{\mailto{hector.gisbert@tu-dortmund.de, victor.miralles@roma1.infn.it, joan.ruiz@ific.uv.es}}}

\affil[a]{Fakultät für Physik, TU Dortmund, Otto-Hahn-Str.\,4, D-44221 Dortmund, Germany}
\affil[b]{INFN, Sezione di Roma, Piazzale A. Moro 2, I-00185 Roma, Italy}
\affil[c]{IFIC, Universitat de València-CSIC, Apt. Correus 22085, E-46071 València, Spain }

\date{}

\usepackage[dvipsnames]{xcolor}

\begin{document}

\maketitle

\vspace*{-1cm}

~
\begin{abstract}
	New coloured scalars in the Manohar-Wise model give sizeable contributions to the Electric Dipole Moment (EDM) of the neutron through one-loop and two-loop diagrams, computed in Reference~\cite{Gisbert:2021htg}. Contributions from neutral scalars cancel out for degenerate values of the masses, whose difference is related to the oblique parameters $S$ and $T$.
	The Manohar-Wise model is able to explain the recent value of the $W$-boson mass published by the CDF collaboration. We show, however, that the case of total degeneracy of masses is strongly disfavoured 
	in this model, whose parameter space is substantially reduced when considering also unitarity bounds.
\end{abstract}

\begin{figure}[h!]
	\centering
	\vbox{
		\includegraphics[scale=0.9]{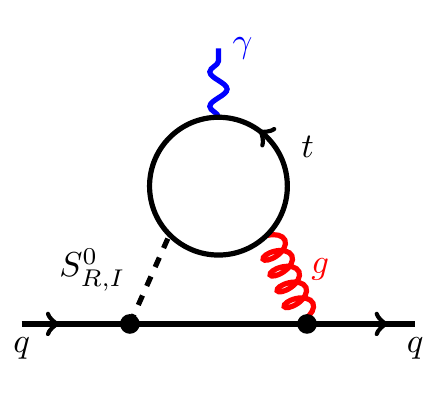}
		\includegraphics[scale=0.9]{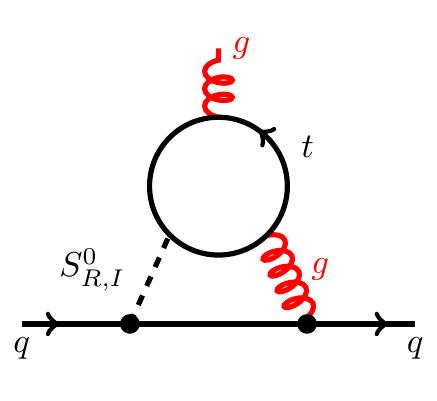}
		\raisebox{0.4cm}{\includegraphics[scale=0.85]{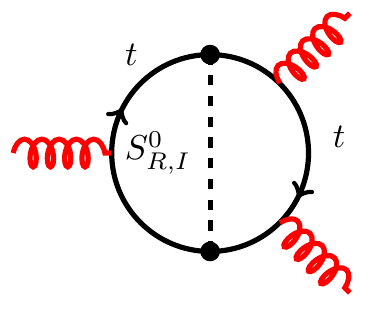}
		\includegraphics[scale=0.85]{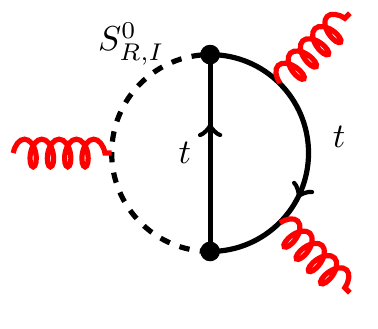}}
	}
	\caption{Barr-Zee type diagrams contributing to the quark EDM and chromo-EDM; and leading contributions to the Weinberg operator in the MW model~\cite{Gisbert:2021htg}. 
	}\label{fig:barr_zee}
\end{figure}

The Manohar-Wise (MW) model introduces new colour-octet scalar fields, which are electroweak doublets. The model was initially motivated as a possibility to implement the principle of Minimal Flavor Violation (MFV)~\cite{Manohar:2006ga}. In addition, these scalars emerge naturally with a mass of few TeVs from $SU(4)$, $SU(5)$ or $SO(10)$ unification theories at high energy (see \textit{e.g.}~\cite{FileviezPerez:2013zmv, Perez:2016qbo,Dorsner:2007fy}). To derive robust limits on the (CP-violating) Yukawa couplings of these scalars, the direct contributions to the neutron EDM through gluonic or light-quark operators need to be considered. The most relevant contributions, in Figure~\ref{fig:barr_zee}, have been computed only recently in Ref.~\cite{Gisbert:2021htg}.

The EDM of heavy quarks carries complementary information to restrict new physics models. Several examples are given in~Ref.~\cite{Gisbert:2019ftm}, where it is shown that its effect is most important in models with non-trivial flavour structure, \textit{e.g} with family-specific free parameters. However, even in models with MFV, like the MW theory, the phenomenological relevance of heavy-quark EDMs should not be overlooked.

\begin{figure}[h!]
	\centering
	\includegraphics[width=0.43\linewidth]{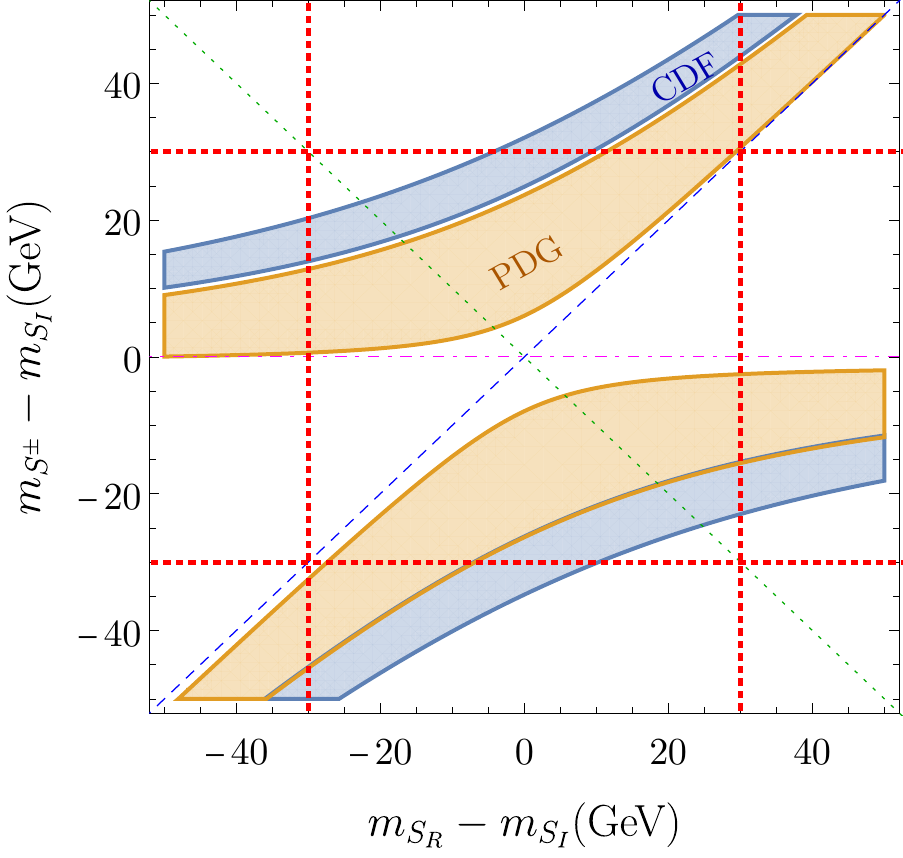}~~
	\includegraphics[width=0.43\linewidth]{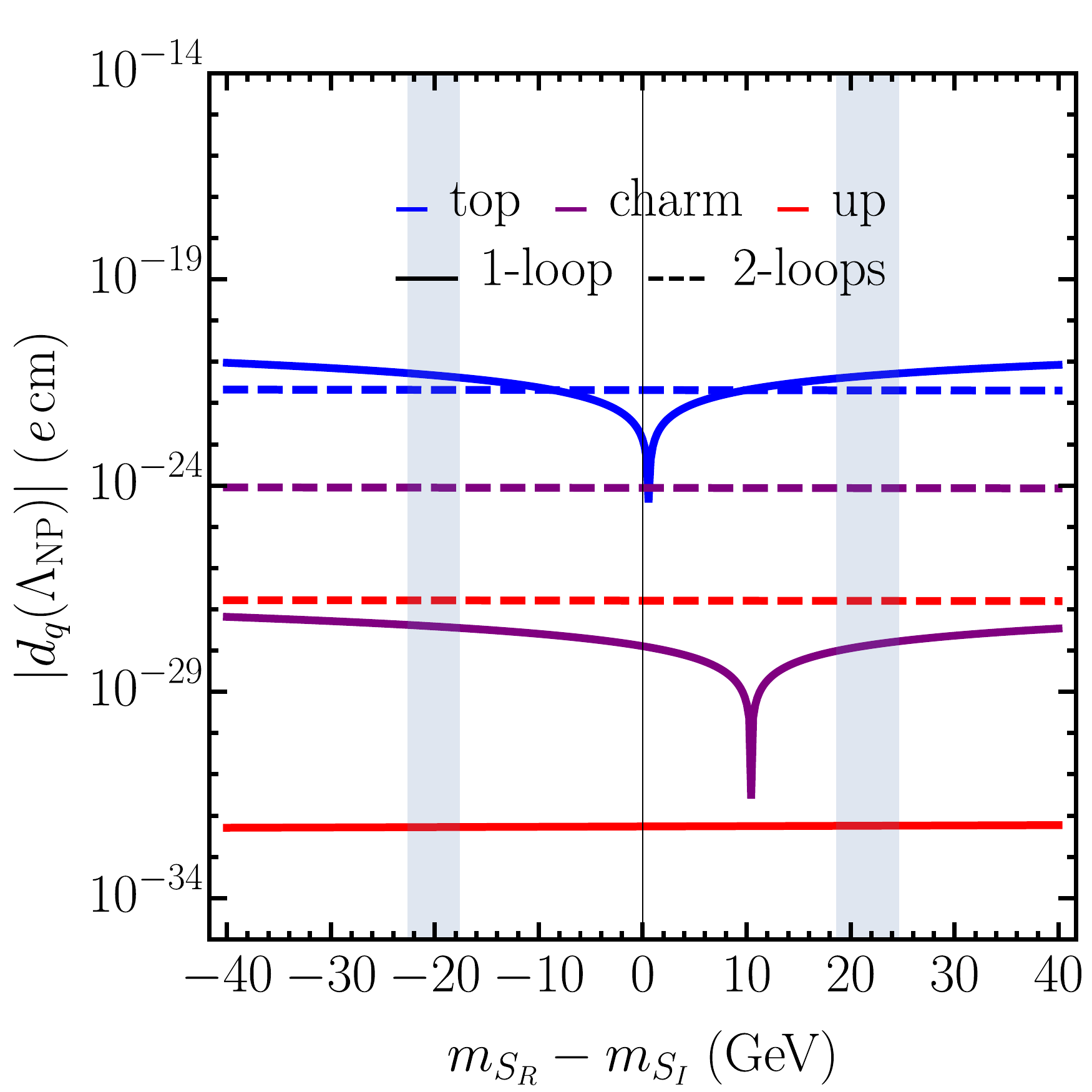}
	\caption{\textbf{Left:} Allowed regions at 1$\sigma$ for the mass splittings of the new scalars imposed by the electroweak-fit values of $S$ and $T$ using the PDG average~\cite{Lu:2022bgw} (orange shaded region), and including the CDF result~\cite{deBlas:2022hdk} (blue shaded region). 
	The CDF value of the $W$ mass strongly disfavours (by almost $\sim 6\,\sigma$) a total mass degeneracy at $(0,\,0)$, that is $m_{S^\pm}\approx m_{S_I}\approx m_{S_R}$. Unitarity bounds on the MW model, on the other hand, impose ${|m_{S_i} - m_{S_j}|\lesssim30~\text{GeV}}$ (dotted red lines)~\cite{Eberhardt:2021ebh}. In this plot, $m_{S_I} \sim m_S
	\sim 1~\text{TeV}$~\cite{Miralles:2019uzg}. 
	\textbf{Right:}  EDMs of the up-type quarks as a function of the neutral-scalar mass splitting. The gray bands, ${m_{S_R}-m_{S_I}=-20.1\pm 2.5\,\text{GeV}}$ and ${-21.6\pm 3.0\,\text{GeV}}$, represent the preferred values by the CDF $W$ mass measurement when the two mass splittings are assumed to be related as $ (m_{S_R} - m_{S_I}) = \beta \,(m_{S^\pm} - m_{S_I})$ with $\beta=-1$ (left plot, green line). Some other $\beta$ configurations, such as $\beta=0$ and $\beta=1$ (magenta and blue lines, respectively) have no overlap with the blue region, \textit{i.e.} are not compatible with the CDF measurement. In this (right) plot, the Yukawa couplings have been fixed to $|\eta_U| = |\eta_D|= 1$, $ \arg(\eta_U)=\pi/4$, and $ \arg(\eta_D)=0$; and the CP-odd scalar mass to ${m_{S_I} = 1~\text{TeV}}$. 
	}
	\label{fig:EDM_heavy_quarks}
\end{figure}

It is interesting to compare the size of one- and two-loop contributions across the quark flavours. If the Yukawa couplings of the new scalars are proportional to the quark masses, the light-quark EDMs are typically dominated by two-loop Barr-Zee contributions, which are enhanced by the top-quark mass~\cite{Gisbert:2021htg}. With the same argument, the EDM of heavy quarks should be dominated by one-loop contributions. In the MW model, this is satisfied for down quarks, as $d_b^{\rm 1-loop} > d_b^{\rm 2-loop}$, $d_s^{\rm 1-loop} < d_s^{\rm 2-loop}$ and $d_d^{\rm 1-loop} < d_d^{\rm 2-loop}$. However, for reasonable values of the model parameters, the top-quark EDM appears to be also dominated by the two-loop Barr-Zee diagram. The reason can be found by inspecting the expressions at one-loop level. For $m_{S_R} = m_{S_I}$, the neutral scalar contributions cancel out at one-loop level, and the charged contribution is suppressed by the mass of the bottom quark. In contrast, the Barr-Zee contribution only contains top-quark masses. The dependence of the quark EDM with the mass splitting is shown in Figure~\ref{fig:EDM_heavy_quarks} (right), where we see that the top-quark one-loop contribution is strongly affected (note the logarithmic scale).

After the conference \textsc{lepton-photon 2021}, but before the publication of these proceedings, the CDF collaboration published a new measurement of the $W$-boson mass in tension with the SM prediction~\cite{CDF:2022hxs},
\begin{equation}
	m_{W^\pm}^{\rm CDF} =  80.4335 \pm 0.0094\,\text{GeV}.
\end{equation}
The MW model can account for this anomalous value of the $W$ mass via loop corrections. In turn, accommodating this result has direct consequences on the mass splitting of the new scalars which, as we have seen, determine the hierarchy of quark EDM contributions. In the following we will briefly explore some consequences of the CDF measurement on the MW model, complementing the published studies on other scalar extensions of the SM~\cite{Ghorbani:2022vtv,Ahn:2022xeq,Song:2022xts,Perez:2022uil,Han:2022juu,Kanemura:2022ahw,Heo:2022dey,Babu:2022pdn,DiLuzio:2022xns,Bahl:2022xzi,Carpenter:2022oyg,Lee:2022gyf,Fan:2022dck,Addazi:2022fbj,Popov:2022ldh,Miralles:2022jnv}.

Electroweak global fits taking the new world-average of the $W$ mass find new minima for the Peskin-Takeuchi oblique parameters $S,~T,~U,$~\cite{deBlas:2022hdk,Lu:2022bgw,Asadi:2022xiy}. Notably, the $U$ parameter gets a larger value than $S$ and $T$, while the opposite is expected from the dimensionality  of the couplings involved.
For this reason, we consider the case $U=0$ and take the $S$ and $T$ values fitted with the PDG average of the $W$ mass $(S=0.05\pm0.08,~T=0.09\pm0.07)$~\cite{Lu:2022bgw}, and the CDF result $(S=0.100\pm0.073 ,~T=0.202\pm0.056)$~\cite{deBlas:2022hdk} (find correlations in the original references).

In the MW model, these parameters are specially sensitive to the mass difference between the new scalar degrees of freedom, with masses $m_{S^\pm}$, $m_{S_R}$, and $m_{S_I}$, while $S$ is also sensitive to the non-physical parameter $m_S$, representing the mass of the unbroken scalar doublet~\cite{Manohar:2006ga}. The expressions for $S$ and $T$ read~\cite{Burgess:2009wm}
\begin{align}\label{eq:ST}
    T \approx -\frac{2\,\delta m_{\pm I}\,\delta m_{RI}}{3\,\pi\,\text{sin}^2\theta_W\,\text{cos}^2\theta_W\,M_{Z^0}^2}\,,\quad S\approx\frac{2\,m_{S_I}\,(\delta m_{RI}\,-\,2\,\delta m_{\pm I})}{3\,\pi\,m_{S}^2}~.
\end{align}
Here the expressions are expanded to leading order on the mass splittings ${\delta m_{ij}\,=\,m_{S_i}-m_{S_j}}$, while the full expression is used later in Figure~\ref{fig:EDM_heavy_quarks}\footnote{A critical reader  could be surprised by the fact that $T$, in Eq.~\eqref{eq:ST}, seems to be independent of the absolute scalar mass. Indirectly, however, the mass splitting is proportional to the parameters of the potential and to the inverse power of the scalar mass.}. 
Compared to the two-higgs-doublet model, the MW theory has an additional colour factor (${\rm dim} = 8$) in the oblique parameters~\cite{Manohar:2006ga}, increasing the sensitivity of these electroweak precision observables to the model parameters.
In Eq.~\eqref{eq:ST}, the Weinberg angle $\theta_W$ is fixed by the electroweak input parameters $G_F$, $m_{Z^0}$, and $\alpha_e$~\cite{Brivio:2021yjb}. 

The allowed region of mass splittings is shown in Figure~\ref{fig:EDM_heavy_quarks} (left), where we see that the case of totally degenerate masses is strongly disfavoured (at $6\,\sigma$ level) with the CDF measurement of the $W$ mass.
In addition to the experimental constraints from the oblique parameters, restrictions on the mass splittings of the colour-octet scalars are also obtained from imposing perturbative unitarity and renormalization-group stability \cite{He:2013tla,Cheng:2018mkc,Eberhardt:2021ebh,Cao:2013wqa}. Indeed, using these theoretical constraints the mass splitting is reduced to be smaller than 30 GeV for masses of the scalars of around 1 TeV, as shown in Ref.~\cite{Eberhardt:2021ebh}.
The combination of the oblique parameters with the theoretical restrictions reduce the parameter space of the theory to the blue regions (oblique parameters) between the red lines (theoretical constraints). In particular, the mass difference $|m_{S^\pm} - m_{S_R}| \approx  [12,30]~\text{GeV}$, for masses of the coloured scalars of around 1 TeV.

As shown in Fig.~\ref{fig:EDM_heavy_quarks} (right), taking the benchmark
$ (m_{S_R} - m_{S_I}) = -(m_{S^\pm} - m_{S_I})$
inspired by the CDF $W$-mass value,
the one-loop contribution of the top-quark EDM is greatly enhanced and dominates over the Barr-Zee diagrams (see caption of Fig.~\ref{fig:EDM_heavy_quarks}). This enhanced value of the top-quark EDM lies close to its experimental bound, $|d_t| \lesssim 10^{-20}\,e\,\text{cm}$~\cite{Cirigliano:2016njn}, and the phenomenological consequences of this observable together with the $W$ mass should be studied further within this model.

\newpage

\footnotesize{
\bibliographystyle{utphys}
\bibliography{references.bib}
}

\end{document}